\documentclass[prl,twocolumn,amsmath,superscriptaddress,amssymb]{revtex4}

\usepackage{graphicx}
\usepackage{dcolumn}
\usepackage{bm}
\usepackage{amssymb}
\usepackage{amsmath}
\usepackage{wasysym}
\usepackage{color}
\usepackage{times}

\usepackage{placeins}
\usepackage{epstopdf}
\usepackage[ngerman,english]{babel} % CH Min added

\newcommand{\rv} {{\mathbf r}}

\newcommand{\yu} {{\chi^{\uparrow}  }}
\newcommand{\yd} {{\chi^{\downarrow}}}

\usepackage[normalem]{ulem}
\usepackage{natbib} % CH Min added
\makeatletter
\renewcommand*\env@matrix[1][c]{\hskip -\arraycolsep
  \let\@ifnextchar\new@ifnextchar
  \array{*\c@MaxMatrixCols #1}}
\makeatother
\usepackage{natbib}
\begin{document}

\title{Orbital Fingerprint of Topological Fermi Arcs in the Weyl Semimetal TaP}
\author{Chul-Hee Min}\affiliation{Experimentelle Physik VII, Universit\"at W\"urzburg, Am Hubland, D-97074 W\"urzburg, Germany}
\author{Hendrik Bentmann}\email{Hendrik.Bentmann@physik.uni-wuerzburg.de}\affiliation{Experimentelle Physik VII, Universit\"at W\"urzburg, Am Hubland, D-97074 W\"urzburg, Germany}
\author{Jennifer N. Neu}\affiliation{National High Magnetic Field Laboratory, Tallahassee, FL 32310, USA}
\author{Philipp Eck}\affiliation{Theoretische Physik I, Universit\"at W\"urzburg, Am Hubland, D-97074 W\"urzburg, Germany}
\author{Simon K. Moser}\affiliation{Advanced Light Source, Lawrence Berkeley National Laboratory, Berkeley, CA 94720, USA}
\author{Tim Figgemeier}\affiliation{Experimentelle Physik VII, Universit\"at W\"urzburg, Am Hubland, D-97074 W\"urzburg, Germany}
\author{Maximilian \"Unzelmann}\affiliation{Experimentelle Physik VII, Universit\"at W\"urzburg, Am Hubland, D-97074 W\"urzburg, Germany}
\author{Katharina Treiber}\affiliation{Experimentelle Physik VII, Universit\"at W\"urzburg, Am Hubland, D-97074 W\"urzburg, Germany}
\author{Peter Lutz}\affiliation{Experimentelle Physik VII, Universit\"at W\"urzburg, Am Hubland, D-97074 W\"urzburg, Germany}
\author{Roland Koch}\affiliation{Advanced Light Source, Lawrence Berkeley National Laboratory, Berkeley, CA 94720, USA}
\author{Chris Jozwiak}\affiliation{Advanced Light Source, Lawrence Berkeley National Laboratory, Berkeley, CA 94720, USA}
\author{Aaron Bostwick}\affiliation{Advanced Light Source, Lawrence Berkeley National Laboratory, Berkeley, CA 94720, USA}
\author{Eli Rotenberg}\affiliation{Advanced Light Source, Lawrence Berkeley National Laboratory, Berkeley, CA 94720, USA}
\author{Ronny Thomale}\affiliation{Theoretische Physik I, Universit\"at W\"urzburg, Am Hubland, D-97074 W\"urzburg, Germany}
\author{Giorgio Sangiovanni}\affiliation{Theoretische Physik I, Universit\"at W\"urzburg, Am Hubland, D-97074 W\"urzburg, Germany}
\author{Theo Siegrist}
\affiliation{Department of Chemical and Biomedical Engineering, FAMU-FSU College of Engineering
   Tallahassee, FL 32310, USA}\affiliation{National High Magnetic Field Laboratory, Tallahassee, FL 32310, USA}
\author{Domenico Di Sante}\affiliation{Theoretische Physik I, Universit\"at W\"urzburg, Am Hubland, D-97074 W\"urzburg, Germany}
\author{Friedrich Reinert}\affiliation{Experimentelle Physik VII, Universit\"at W\"urzburg, Am Hubland, D-97074 W\"urzburg, Germany}

\date{\today}
\begin{abstract}
 
The monopnictides TaAs and TaP are well-established Weyl semimetals. Yet, a precise assignment of Fermi arcs, accomodating the predicted chiral charge of the bulk Weyl points, has been difficult in these systems, and the topological character of different surface features in the Fermi surface is not fully understood. Here, employing a joint analysis from linear dichroism in angle-resolved photoemission and first-principles calculations, we unveil the orbital texture on the full Fermi surface of TaP(001). We observe pronounced switches in the orbital texture at the projected Weyl nodes, and show how they facilitate a topological classification of the surface band structure. Our findings establish a critical role of the orbital degrees of freedom in mediating the surface-bulk connectivity in Weyl semimetals.  
\end{abstract}
\maketitle

The discovery of topological insulators gave birth to the field of topological quantum states in crystalline solids \cite{Hasan:10.11}. In a recent breakthrough, the notion of topological phases of matter was extended from insulators to gapless semimetals \cite{Wan:11,Burkov:11,Bradlyn:16,Armitage:18}. A paradigmatic example is the Weyl semimetal \cite{Wan:11,Burkov:11,huang:15,Weng:15,Xu:15,Lv:15,Turner:13,Sun:15,inoue:16,Armitage:18}, whose band structure features topologically protected band touching points near the Fermi level, called Weyl nodes. Probably the most striking manifestation of the topological properties in a Weyl semimetal is the existence of unconventional Fermi-arc surface states. As a function of surface momentum, these states form non-closed Fermi-surface segments that connect to the surface projections of a pair of bulk Weyl nodes of opposite chirality \cite{Wan:11,Burkov:11}. 

The monopnictides TaAs \cite{huang:15,Xu:15,Lv:15,Armitage:18} and TaP \cite{Sun:15,nxu:16,Xu_TaP:15,liu:16} are well-established and paradigmatic Weyl semimetals. Nevertheless, previous measurements of the surface electronic structure by angle-resolved photoemission (ARPES) in TaP(001) showed substantial differences \cite{Xu_TaP:15,liu:16} and, thus, do not provide a consistent picture of the surface-bulk connectivity. Similarly, in TaAs(001) the assignment of measured surface states as topologically trivial and non-trivial has been inconsistent \cite{Xu:15,Lv:15}. Even elaborate calculations based on density functional theory (DFT) could not conclusively clarify the topological connection of the surface states to the bulk band structure \cite{Sun:15}. Thus, central aspects of the surface-bulk connectivity in these materials remain unclear. 

It has been proposed that the spin texture of the Fermi arcs could provide a tag to analyze their connectivity to the bulk bands \cite{Sun:15}, which, however, is experimentally challenging to achieve with sufficient momentum resolution. On the other hand, previous work on topological insulators showed the importance of orbital textures for a realistic description of the topological surface states and a variety of their properties \cite{Hasan:10.11,Kim:12,cao:13,zeljkovic:14,xie:14}. Here, we use ARPES experiments and first principles calculations to determine the orbital texture of the surface Fermi-surface in TaP(001). We observe pronounced orbital-symmetry changes at the projected bulk Weyl points. According to our calculations, these changes reflect transitions from genuine surface states to surface resonances with large bulk delocalization and bulk-like orbital character. These subtle effects are not captured by effective models \cite{Turner:13,Armitage:18}, but they turn out to be critical to reconcile the observed surface band structure with the calculated chiral charge of the Weyl nodes \cite{huang:15,Sun:15}. 

\begin{figure*}[t]
\includegraphics[width=\textwidth]{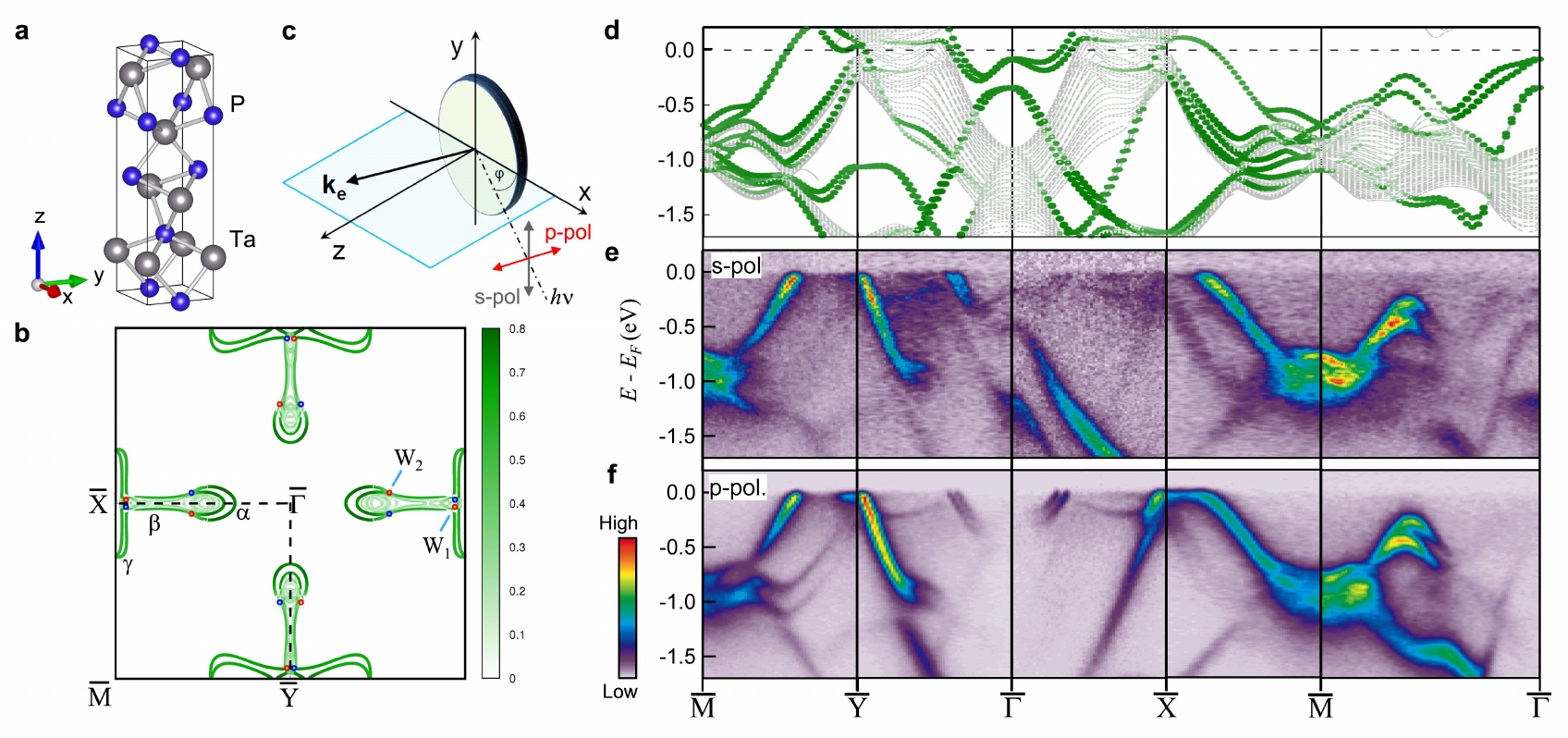}
\caption{(a) Bulk crystal structure of TaP with space group $I4_1 md$. (b) Calculated Fermi surface of the P-terminated TaP(001) surface with $C_{2v}$ symmetry. The features $\alpha$, $\beta$ and $\gamma$ as well as the  positions of the projected Weyl points W$_1$ and W$_2$ are indicated. (c) Experimental geometry of the angle-resolved photoemission (ARPES) experiment with linearly polarized light. (d)-(f) Band structure of P-terminated TaP(001) along high-symmetry directions, as determined by a first-principles calculation and by ARPES using $s$- and $p$-polarized light at $h\nu =$~105~eV. In (d) the green dots refer to electronic states on the P-terminated surface. Dark (light) green refers to a high (low) surface character. The calculated bands were shifted upwards rigidly by 50~meV to obtain the best match with the experimental data.  
}
\label{fig1}
\end{figure*}

We performed ARPES experiments on TaP single crystals \cite{Supp1,Li:16,Rigaku:16}~at the MAESTRO endstation at beamline 7 of the Advanced Light Source (ALS) in the photon energy range of 60 eV $-$180 eV. Measurements were performed in ultra-high vacuum of lower than 1~$\times$ 10$^{-10}$ mbar for the samples cooled down below 20~K. The slit of a Scienta R4000 electron analyzer was aligned in the plane of light incidence, namely the $xz$-plane as shown in Fig.~\ref{fig1}(c). For the theoretical study of TaP we considered a supercell of 7 unit cells along the (001) direction and applied DFT as implemented in the VASP code \cite{Kresse:96}.
The projector-augmented-plane-wave (PAW) method was used by expanding the Kohn-Sham wave functions into plane waves to an energy cut-off of 300 eV \cite{Kresse:99}. Spin-orbit coupling was included self-consistently and the exchange-correlation was computed within the generalized gradient approximation as parameterized in the PBE-GGA functional \cite{Perdew:96}. Additional experimental and theoretical details are given in the supplementary information.

TaP crystallizes in the non-centrosymmetric space group $I4_1 md$, as shown in Fig.~\ref{fig1}a. Its bulk band structure features 24 Weyl points in the Brillouin zone \cite{huang:15,Weng:15,Xu:15,Lv:15,Sun:15,nxu:16,Xu_TaP:15,liu:16}. A projection on the (001) plane gives rise to two inequivalent sets of projected Weyl points, W$_1$ and W$_2$. The calculated band structure of the P-terminated TaP(001) surface in Fig.~\ref{fig1}d highly accurately matches with our ARPES measurements in Fig.~\ref{fig1}e-f, in particular regarding the bands close to the Fermi level which are relevant for the topological properties. In turn, by comparing theory and experiment, we identify all main spectral features in the experimental data as surface states. The calculated Fermi surface of TaP(001) in Fig.~\ref{fig1}b is composed of the spoon-shaped feature $\alpha$, the neck feature $\beta$, as well as the outer bowtie-shaped feature $\gamma$ around the $\bar{X}$- and $\bar{Y}$-points. As seen in Fig.~\ref{fig1}b, the projected Weyl nodes W$_2$ and W$_1$ are located approximately at the connection points of the features $\alpha$ and $\beta$ as well as $\gamma$ and $\beta$, respectively.

\begin{figure*}[t]
\includegraphics[width=6in]{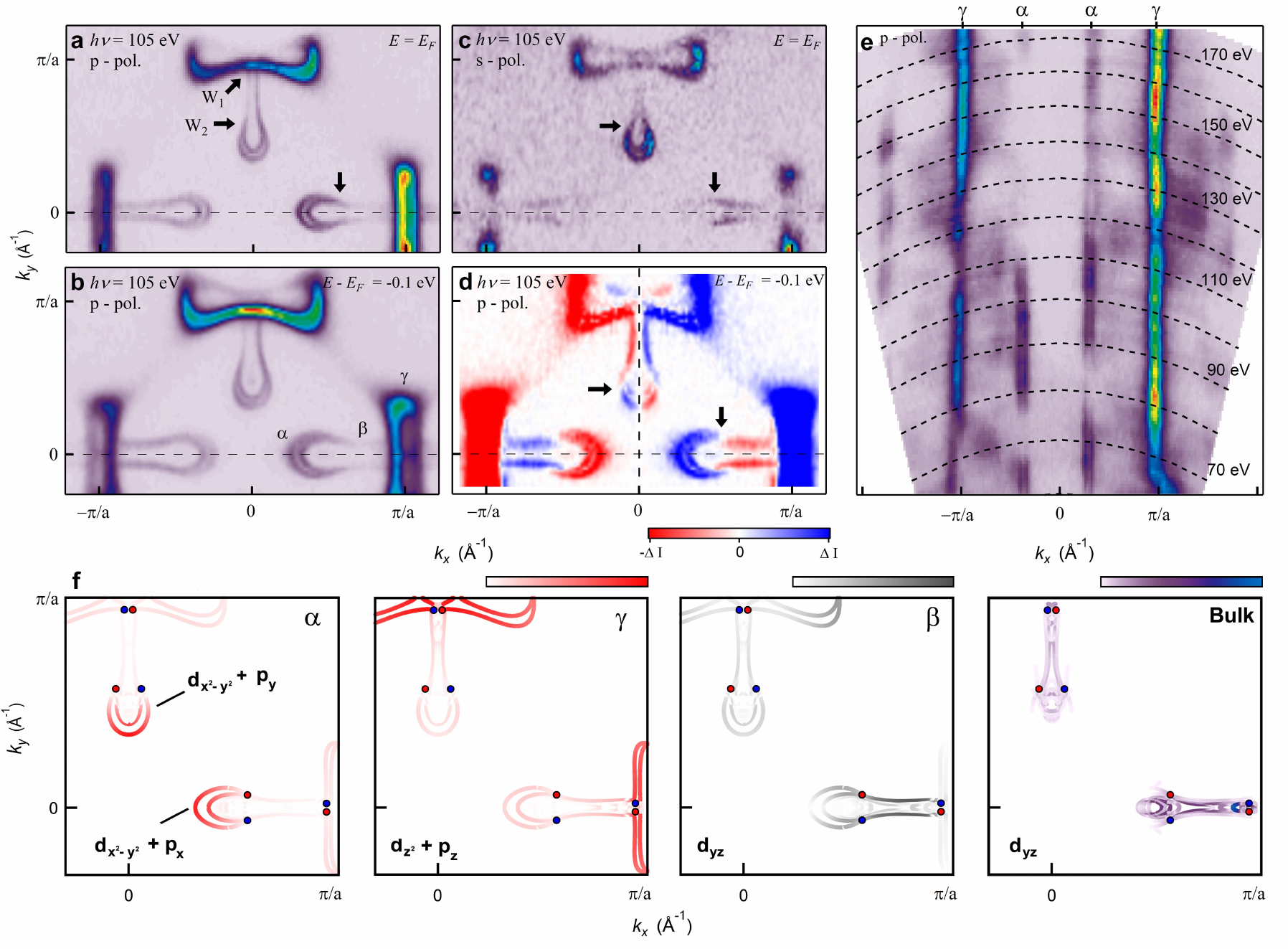}
\caption{(a)-(c), ARPES constant-energy-contours of the surface band structure at and near Fermi level $E_F$ obtained with $s$- and $p$-polarized light. (d) Linear dichroism, defined as the intensity asymmetry $\Delta I = I(k_x,k_y)-I(-k_x,k_y)$, obtained from the measured intensity $I(k_x,k_y)$ with $p$-polarization in panel (b). Near the projected Weyl points W$_1$ and W$_2$ [cf. Fig~\ref{fig1}b] pronounced changes in the ARPES intensity and in the linear dichroism are observed (indicated by arrows), evidencing pronounced rearrangements in the orbital composition. (e) Photon-energy-dependence of the ARPES intensity at $E_F$ and $k_y =$0 for $p$-polarized light. (f) Calculated Fermi surface projected on different Ta 5$d$ and P 3$p$ orbitals in the surface unit cell. A projection on the $d_{yz}$ orbital is shown also for a unit cell in the bulk. 
}
\label{fig2}
\end{figure*}

Based on the high consistency of the experimental and {theoretical} band structures, we shall now address the orbital composition of the Fermi-arc surface states. To this end we make use of parity selection rules within the dipole approximation for photoexcitation with linearly polarized light \cite{cao:13,xie:14}. The experimental geometry is shown in Fig.~\ref{fig1}c, in which the light electric field vector is given by $(0,\mathcal{E}_y ,0)$ for $s$-polarization and by $(\mathcal{E}_x ,0,\mathcal{E}_z )$ for $p$-polarization. We acquired ARPES data as a function of both in-plane momentum components without changing the experimental geometry by employing custom-made deflectors. Using $p$-polarized light, incident along the emission plane, only even orbital components of the initial state contribute to the photoemission intensity, while for $s$-polarized light only odd components contribute. As indicated by arrows in Fig.~\ref{fig2}a-c, for different light polarizatons we observe pronounced intensity variations in the measured Fermi-surface contours close to the projected bulk Weyl points W$_1$ and W$_2$ [cf. Fig.~\ref{fig1}b]. 
 
For $s$-polarized light the intensity of the states $\alpha$ and $\gamma$ is almost entirely suppressed along $k_x$, i.e. within the plane of light incidence (Fig.~\ref{fig2}c and Fig.~S2). This suppression persists over a wide range of excitation energies (see supplementary note I), demonstrating that it is independent of the photoelectron final state and related to the symmetry of the initial state. From the experimental data, we therefore infer an almost purely even orbital symmetry for the states $\alpha$ and $\gamma$ with respect to the $xz$ mirror plane. This pure orbital symmetry of the Fermi-arcs is in striking contrast to the orbital wave-function of the Dirac states in topological insulators \cite{cao:13,xie:14}.

Our first-principles calculations in Fig.~\ref{fig2} confirm the experimentally determined orbital symmetry. Along $k_x$ the state $\alpha$ is composed predominantly of Ta $d_{x^{2}-y^{2}}$ and P $p_x$ orbitals, and the state $\gamma$ consists mainly of P $p_z$ and Ta $d_{z^{2}}$ out-of-plane orbitals.\cite{Xu:16} Hence, while both features derive from even orbitals, the overall orbital texture on the Fermi surface varies from out-of-plane near the Brillouin zone boundary towards in-plane near the center. This is corroborated by the photon-energy-dependent ARPES data in Fig.~\ref{fig2}e obtained with $p$-polarized light: Along either $+ k_x$ or $- k_x$, the state $\alpha$ shows complete intensity suppressions over several intervals of photon energy. Such a pronounced $h\nu$-dependent linear dichroism is not observed for $\gamma$, which can be attributed to the different orbital composition of the two states \cite{Henk:04,Bentmann:17}, as detailed in supplementary note II.

Considering next the full momentum dependence of the ARPES data sets, we find that the intensity distributions obtained with $s$-polarized light show sharp changes at wave vectors, where the spoon feature $\alpha$ evolves into the neck feature $\beta$ (Fig.~\ref{fig2}c and Fig.~S2). These wave vectors match reasonably well the positions of the projected W$_2$ Weyl points, reported to be close to the Fermi level and at ($\pm$0.54,$\pm$0.03)$\frac{\pi}{a}$ \cite{nxu:16}. Along $k_x$ the intensity of $\alpha$ is suppressed and $\beta$ shows appreciable intensity, while along $k_y$ we observe the opposite behaviour. Our calculations in Fig.~\ref{fig2}f demonstrate how this is related to a rapid change of the Fermi-arc orbital texture near W$_2$ from even $d_{x^{2}-y^{2}}$ and $p_x$ character to mainly odd $d_{yz}$ character (see also supplementary notes I and III). 

Similarly pronounced intensity variations at the projected Weyl points W$_1$ and W$_2$ are observed when using $p$-polarized light in Fig.~\ref{fig2}a-b. For $p$-polarization it is particularly useful to consider the linear dichroism, which can be defined as the intensity asymmetry $I(k_x,k_y)-I(-k_x,k_y)$ (see supplementary note II). The linear dichroism sensitively traces orbital-character changes as a function of momentum \cite{Henk:04,Chernov:15,Tusche:16,Schonhense:17,Bentmann:17}. Indeed, the measured linear dichroism in Fig.~\ref{fig2}d shows pronounced changes, even sign reversals evident from red/blue switches, at the projected Weyl points. Note that the linear dichroism varies with photon energy as seen in Fig.~\ref{fig2}e, and, therefore, does not directly reflect a specific orbital. Rather, it is the abrupt reversal with momentum that reflects the sudden change in orbital character which is predicted by our calculations and also observed in the measurements with s-polarized light.

\begin{figure*}[t]
\includegraphics[width=6in]{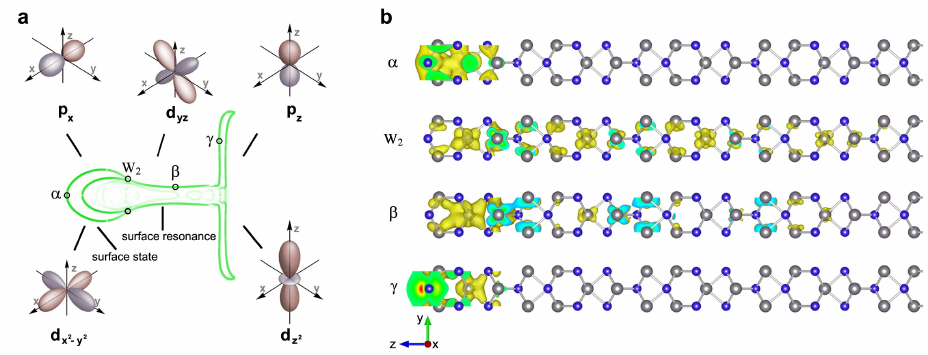}
\caption{(a) Calculated Fermi surface with schematic indications of the varying orbital character deduced from the ARPES experiments and first-principles calculations in Fig.~\ref{fig2}. (b) Partial charge densities of the Fermi arcs calculated at four selected positions on the Fermi surface indicated in (a). The states $\alpha$ and $\gamma$ are highly localized in the first few atomic layers of the surface and therefore can be classified as genuine surface states. By contrast, near the W$_2$ Weyl point and for the state $\beta$ the wave function propagates deep into the bulk, indicating a surface-resonance character.   
}
\label{fig3}
\end{figure*} 
 
Our calculations in Fig.~\ref{fig3} show how the changes in orbital character are linked to the momentum-dependent bulk propagation of the Fermi arcs. The states $\alpha$ and $\gamma$ are well-separated from the projected bulk Fermi surface and accordingly their wave functions are highly surface-localized \cite{inoue:16}. Therefore, we classify $\alpha$ and $\gamma$ as genuine surface states. Instead, near the projected Weyl points and for the feature $\beta$ the wave functions propagate deeper into the bulk, as seen in Fig.~\ref{fig3}b. Furthermore, the orbital character of $\beta$ is adapted to the nearby bulk continuum states (Fig.~\ref{fig2}f). We therefore classify $\beta$ as a surface resonance with more bulk-like properties than $\alpha$ \cite{Cohen:78}.

The projected Weyl points W$_2$ carry a chirality of $\pm 2$ \cite{huang:15}. Therefore, one expects that two Fermi arcs terminate at W$_2$. However, both, in experiment and theory three surface features meet at W$_2$, namely $\beta$ as well as the two spin-split states $\alpha$ \cite{Xu:15,Lv:15,Sun:15,liu:16}. Previous works for TaAs either classified the two $\alpha$ states\cite{Xu:15} or the outer $\alpha$ state and the feature $\beta$ \cite{Lv:15} as the two Fermi arcs terminating at W$_2$. A more detailed discussion of the band connectivity at W$_2$ was presented in Ref.~\cite{Belopolski:16}. However, the analysis did not consider the presence of $\beta$ and its role in the band connectivity. Based on calculations for TaP the outer $\alpha$ state was identified as a Fermi arc \cite{Sun:15}. However, the character of the two other features was not specified and the question how to reconcile the surface band connectivity with a Weyl point chirality of $\pm 2$ was not addressed and apparently remains open.  

The orbital texture provides an improved perspective on the complex band connectivity: at W$_2$ the surface bands show a sudden change in orbital symmetry, manifesting itself, e.g., in the red-blue switch of the linear dichroism at W$_2$ (Fig.~\ref{fig2}d). This reflects a transition from the true surface state $\alpha$ to the more bulk-delocalized surface resonance $\beta$. We propose that it is precisely this transition that characterizes the bulk connectivity of the Fermi arcs $\alpha$, which extend as surface resonance $\beta$ into the projected bulk continuum. This penetration of the surface bands $\alpha$ into the projected bulk states is also seen in the calculated band structure in Fig.~\ref{fig1}d. 

While the proposed scenario reconciles the expected connectivity of two genuine surface bands at W$_2$, it deviates from idealized models in which the Fermi-arc wave function fully delocalizes into the bulk at the projected Weyl points \cite{Turner:13}. These deviations in comparison to model calculations are related to the more complex bulk Fermi surface in TaP and TaAs, in which additional electron and hole pockets coexist with the Weyl points close to $E_F$ \cite{Sun:15}. Our considerations show that in such cases a mere classification into pure surface and bulk states might not be sufficient for an analysis of the topological band connectivity. Instead our findings highlight the importance to distinguish between genuine surface states and more bulk-like surface resonances. As we show, this distinction may be achieved on the basis of the measured and calculated orbital texture. We note that the importance of surface resonances has also been discussed in the surface-bulk band connectivity in topological insulators \cite{Seibel:15,jozwiak:16}.  

We explore implications of the determined orbital symmetries for the spin polarization of the Fermi arcs $\alpha$, which are split by spin-orbit interaction into two branches as verified by calculations and spin-resolved ARPES \cite{Lv:15_2,Xu:16,Sun:15}. Along the $xz$ mirror plane we can write the surface-state wave functions as $\Psi=g\yu + u\yd$, with $\chi^{\sigma}$ quantized along $y$ \cite{xie:14,Bentmann:17}. The spinor component $g(\rv)$ is even under mirror reflection $y\to-y$, and $u(\rv)$ is odd. Remarkably, our experiments demonstrate that for the Fermi arcs $\alpha$ the odd contribution $u(\rv)$ is negligible within experimental uncertainty, which implies a nearly full spin polarization (see also supplemenatry note IV). In good agreement, our calculations find a spin polarization of $P\sim 0.96$ for the outer $\alpha$ state along $k_x$, as well as a ratio of only 2~\% between odd and even orbitals contributing to $\Psi$. Spin-resolved photoemission experiments showed values between ca.~0.3-0.8 depending on photon energy \cite{Lv:15_2,Xu:16}, indicating possible final-state effects. The spin polarization of the Fermi arcs in TaP is therefore much higher than for the Dirac states in typical topological insulators with only $P\sim 0.5$ \cite{Yazyev:10}, which might be relevant for spintronic applications.  

According to our calculations the spin polarization of $\alpha$ remains high also for wave vectors off the mirror plane despite an increasing admixture of other orbital components. Interestingly, however, near the W$_2$ points the spins of the different orbital components develop a considerable non-collinearity, which leads to a reduction of the spin polarization at W$_2$ \cite{Supp2,Tusche:15,maass:16}. We attribute these spin-texture modulations at the W$_2$ points to the strong changes in orbital composition at these wave vectors.

While orbital-related phenomena in the electronic structure of topological and strongly correlated materials have been extensively discussed recently \cite{Kim:12,cao:13,zeljkovic:14,xie:14,wiesner:14,ritschel:15,borisenko:16}, the orbital-texture characteristics of topological Fermi arcs, reported here, are distinct in that they reflect the connectivity of surface and bulk band structure. The latter is a hallmark feature of topological semimetals and underlies unusual electron transport phenomena in these materials \cite{Potter:14,moll:16}. By establishing the importance of orbital textures in topological semimetals our results complement previous findings of orbital-textured Dirac states in topological insulators \cite{cao:13} and topological crystalline insulators \cite{zeljkovic:14}.

\section{Acknowledgments}
H.B. would like to thank Jan Minar for helpful discussions. This work was supported by the DFG through SFB1170 'Tocotronics', RE 1469/13-1, and through SPP-1666, and by ERC-StG-336012-Thomale-TOPOLECTRICS. We gratefully acknowledge the Gauss Centre for Supercomputing e.V. (www.gauss-centre.eu) for funding this project by providing computing time on the GCS Supercomputer SuperMUC at Leibniz Supercomputing Centre (www.lrz.de). J.N and T.S. acknowledge support from the National Research Foundation, under grant NSF DMR-1606952.
The crystal synthesis and characterization was carried out at the National High Magnetic Field Laboratory, which is supported by the National Science Foundation, Division of Materials Research under grants DMR-1157490 and DMR-1644779, the state of Florida and Florida State University. S.M. acknowledges support by the Swiss National Science Foundation (grant no. P300P2-171221). This research used resources of the Advanced Light Source, which is a US Department of Energy Office of Science User Facility under contract no. DE-AC02-05CH11231.\\

%\section{ADDITIONAL INFORMATION} % CH Min added
%The authors declare no competing financial interests. Correspondence and requests for materials should be addressed to Hendrik Bentmann
%(e-mail: hendrik.bentmann@physik.uni-wuerzburg.de).

\end{document}